# Amino acid-driven hydrophobization of alumina and production of particle stabilised foams in a broader pH range

T. dos Santos Junior[(1,*)], C. I. Pereira[(1)], R. Gonçalves[(2)], C. H. M. Fernandes[(3)], V. R. Salvini[(4)], V. C. Pandolfelli[(1)]

(1) Graduate Program in Materials Science and Engineering (PPGCEM), Materials Microstructure Engineering Group (GEMM), FIRE Associate Laboratory, Materials Engineering Department, Federal University of São Carlos, Rodovia Washington Luis, km 235, São Carlos, SP, 13565-905, Brazil

(2) Graduate Program in Materials Science and Engineering (PPGCEM), Smart Materials Research Team (SMaRT), Materials Engineering Department, Federal University of São Carlos, Rodovia Washington Luis, km 235, São Carlos, SP, 13566-905, Brazil

(3) Laboratory of Research Eletrochemistry (LaPE), Chemistry Department, Federal University of São Carlos, Rodovia Washington Luis, km 235, São Carlos, SP, 13566-905, Brazil

(4) College of Technology (FATEC Sertãozinho), Jordão Borghetti Street 480, Sertãozinho, SP, 14160-050, Brazil

*Corresponding author at +55 16 3351 8253 or tiagosjunior@ppgcem.ufscar.br

## Abstract

The interaction of amino acids (glycine, L-valine, L-isoleucine and L-leucine) with alumina surface was studied to induce partial hydrophobization and production of ultrastable particle-stabilised foams. The evaluation of these amino acids was carried out by mechano-quantum simulations followed by experimental tests (foamability, zeta potential, contact angle and foam lifetime measurements). The experimental results agreed with the trends pointed out by the simulations. The selected amino acids interact with alumina particles in aqueous media and in a broader pH range, leading to hydrophobization of surfaces, which was more intense for amino acids with higher molecular mass (isoleucine and leucine). As a consequence, ultrastable foams with a longer lifetime (> 100 hours) were produced and the foam microstructure was preserved from ageing phenomena. Moreover, the attained foams were stable at different pH, opening up new possibilities to develop macroporous multiphasic ceramics, which can result in novel materials for thermal insulation at high temperatures.

Keywords: Amino acids; ultrastable foam; $Al_2O_3$; macroporous ceramic

1. **Introduction**

Foams are systems formed by compressed gas bubbles, originating from a jamming transition, which gives a solid behaviour to the whole, *i.e.* a yield stress [1–3]. These systems hold an important place in nature and imaginary. As pointed out by Ashby [4], natural materials used in constructions, such as wood, bone, nests and shells, are mostly formed by cellular structures. Foams can be found in different peoples´ mythology around the world, *e.g.* in the hindu tale of Indra and Namuchi, described in the Mahabharata, at the birth of the Greek goddess Aphrodite (which literally means "risen from the foams"), or in the genesis of foreign men in the Yanomani mythology [5,6]. Despite human fascination with foam properties, these systems overcome the domain of the imaginary, and technological uses take place in different sectors, occasionally inspired by nature. For example, at high temperatures, refractory macroporous ceramic foams (solid materials with a pore size above 50 nm [7]) can be used as thermal insulators, playing a very important role in the energy conservation of industrial processes. The development of such materials requires the production of liquid foams with good stability, which is a very relevant property to the production of reproductible ceramic foams with low thermal conductivity.

Refractory macroporous ceramics can be prepared by setting a liquid foam containing ceramic particles dispersed in its structure. This processing route, called direct foaming, is considered versatile as macroporous ceramics with different volumetric densities and compositions can be produced. Additionally, direct foaming enables materials to be manufactured with interconnected or completely closed pores, with narrow pore size distribution [8–10]. The incorporation of gas and production of foams containing ceramic particles can be carried out adopting different methods *e.g.* mechanical stirring [11,12], chemical reactions and phase changes [13,14], cavitation [15–17] or gas injection [18,19]. Although gas can be entrapped in a ceramic suspension by different methods, a liquid foam containing ceramic particles should be obtained. This system will be subjected to changes due to its instability.

The surface energy of a foam, $E_{foam}$, can be estimated directly by multiplying its interfacial energy $\gamma$ and the summation of the interfacial areas of all $n$ bubbles stated as $S_i$ [Eq.(1)] [20]. Due to the high surface area, these systems are thermodynamically unstable and tend to undergo modifications in their microstructures in order to reduce their total interfacial area. This is the strategy used by nature, which actuates reducing the total interfacial area of a foam by means of three ageing phenomena named drainage, coarsening and coalescence.

$$E_{foam} = \gamma \sum_{i=1}^{n} S_i \qquad (1)$$

Drainage consists of separating the liquid and gas phases due to the action of gravity over fluids with different volumetric densities. As drainage progresses, the amount of liquid between bubbles on the top of the foam is reduced, resulting in the approximation of neighbouring cells and changing their shapes from spheres to polyhedrons, which is characteristic of dry foams. On the other hand, bubbles located at the bottom of foams are spherical due to the increase in liquid fraction [21,22]. As the distance of bubbles is reduced, the Young-Laplace pressure, *i.e.* the difference of gas pressure between two neighbouring bubbles of different sizes, may favour the diffusion of gas molecules from smaller bubbles to the bigger ones using Ostwald ripening or von Newman-Mullins mechanisms. With the advance of this phenomenon, named coarsening, the increase in the average bubble size is observed [5,21,23]. Simultaneously, as a result of the balance of attractive and repulsive forces acting in the film between two bubbles, which is expressed by the disjoining pressure (resulted from the sum of van der Waals interactions, electrostatic and non-DLVO forces), the attraction of two liquid-air interfaces can take place, resulting in their rupture and the merger of these bubbles, giving rise to a novel one with a bigger volume [5,24,25]. This phenomenon, known as coalescence, leads to the reduction of the number of bubbles and the increase in the average size of those remaining. The action of the three foam ageing phenomena is simultaneous, however, drainage will increase the action of coarsening and coalescence because of the approximation of bubbles. Potentially, the resting foam will reach a metastable equilibrium. In this case, the microstructure would be preserved up to the time where external perturbations, as temperature and pressure variations or mechanical forces, act over the system [25].

Besides the reduction in the surface area induced by the ageing phenomena, the bubbles in a foam can be stabilised if the interfacial energy in equation (1) is reduced. A likely route to reduce $\gamma$ is by the adsorption of solid particles at the liquid-gas interface, replacing part of its area by more stable surfaces during the foaming process. The stabilisation of fluid cells by solid particles has been practised for more than a century in mineral separation and concentration processes [26,27] and for the production of particle stabilised emulsions known as Pickering emulsions [28]. As stated by Aveyard *et al.* [29], the adsorption energy of a spherical solid particle at the interface between two fluids is defined by the affinity of the solid to each of these fluids. This affinity is expressed by means of the contact angle $\theta$. Besides this, the fluid-fluid interfacial energy $\gamma$, the particle and the bubble or vesicle radii, $R_p$ and $R_b$ respectively, the line tension $\tau$, which actuates in the region where the three phases meet, and the radius of this region, x, also are accounted to this property. Thus, the adsorption energy of a spherical particle at the interface between two fluids can be calculated by equation (2):

$$\Delta_{ads}G = 2\pi\tau x + 2\pi\gamma_{lg}\left[R_P^2(1\pm\cos\alpha)\cos\theta\left(1 - \frac{\tau\cos\beta}{\gamma_{lg}x}\right) - R_B^2(1-\cos\beta)\right] \quad (2)$$

where, in addition to the variables discussed before, the angles α and β correspond to *arcsin(x/$R_b$)* and *arcsin(x/$R_p$)*, in this order.

Equation (2) highlights that the energy needed to detach an adsorbed particle at the interface between two fluids will scale with the particle radius and the contact angle. Adsorption energy values close to, or higher than, $10^3$ $k_b$T should be expected for particles attached to the liquid-gas interface of a bubble, which makes the adsorption process irreversible and likely to produce particle stabilised foams. Considering the air-water system, the affinity of the solid particles with these fluids will define whether a foam or a mist will be formed, in a similar way to oil-in-water or water-in-oil Pickering emulsions [29]. Particle stabilised foams are attained when the contact angle of the solid phase with water is between 0º and 90º, and the adsorption energy of these particles at air-water interface is higher for particles with a higher contact angle, *i.e.* partially hydrophobic particles. Partial hydrophobization of solid particles can be carried out by functionalisation of some regions of their surfaces. Thus, partially hydrophobic ceramic particles will be obtained after interacting with organic amphiphilic molecules, *i.e.*, which present polar and non-polar regions in the same structure. The polar region of these molecules interacts with particle surface regions presenting a partial charge density favourable for the interaction, chemically adsorbing and exposing the non-polar regions to the aqueous media. Thus, a non-polar behaviour is imposed to the particle surface region where the amphiphilic molecule is adsorbed, making it partially hydrophobic. The non-polar region will be larger as more molecules interact with the particle surface, making these particles suitable to stabilise air bubbles.

Foams stabilised with silica particles partially hydrophobized with sodium dodecyl sulfonate (SDS) were firstly reported by Tang *et al.* [30] in 1989. In their work, it was observed that using smaller or more hydrophobic particles resulted in more stable foams. Bubbles stabilised with silica were reported by other authors [31–33], however, the main interest was at the foam stability prior to solidification. Binks *et al.* [32] also demonstrated experimentally the phase inversion caused by the change of silica contact angle with water, pointing out the viability of producing stabilised bubbles or liquid marbles armoured with silica particles by adjusting the contact angle for values lower or superior to 90º.

By selecting different amphiphilic organic molecules, Gonzenbach *et al.* [34,35] reported the production of macroporous ceramics of different mineralogical compositions [*e.g.* $Al_2O_3$, $Ca_3(PO_4)_2$, $ZrO_2$ and $SiO_2$], derived from aqueous ultrastable particle stabilised foams. The selected molecules presented amphiphilic features in the pH ranges evaluated. For example, alumina stabilised foams in acid pH were produced with carboxylic acids as propionic and butyric ones, considering that these molecules present a polar head (in the carboxyl position) and a non-polar tail. The stability of these foams was conditioned to a pH range where the organic molecules present amphiphilic behaviour. Based on this condition, the processing of multiphasic compositions was limited. Besides that, some of the substances proposed are toxic, which motivates the search for new molecules [11].

Despite the limitations pointed out before, associated to the processing of particle stabilised ceramic foams, these systems are of technological interest for the production of better thermal insulators for high temperature applications. In addition to their long stability time range, which preserves the foam microstructure comprised by numerous bubbles with small and homogeneous size distribution, it was possible to increase the mechanical strength of these systems by using refractory binders, as calcium aluminate cement or magnesia [36]. Besides that, *in situ* phases could be formed in these foams, reducing their volumetric density and effective thermal conductivity. Unfortunately, adding binders or other ceramic phases could displace pH from acid to basic, causing destabilisation of foams prepared with carboxylic acids.

In order to produce particle stabilised foams that withstand pH changes, novel organic amphiphilic molecules should be selected. These molecules should keep their polar head and non-polar tail in a wider pH range, used in agreement with the Green Chemistry principles and practices [37]. A class of molecules which could respond to these demands are amino acids. They are organic molecules composed by carboxyl (-COOH) and amine ($-NH_2$) functions and a carbonic chain comprised mainly by C and H, although other elements can be present. These molecules are the basic structural units of proteins and, in addition to the fact that they are not toxic, their structures present different electric charges when exposed to aqueous media at any pH. Generally, amino acids are presented as cations in the acid range, due to amine protonation. When close to neutral pH, both the functional groups present charges: carboxyl is deprotonated and presents a negative charge whereas amine is protonated and assumes a positive charge, giving rise to a structure named zwitterion. For basic pH, only the carboxyl remains deprotonated, and the molecule behaves as an anion. The pH values at which the amino acid molecules change from cation to zwitterion or anion may vary from one amino acid to another [38].

Considering the alumina-amino acid system in aqueous solution at different pH values, it is possible for the molecule to adsorb onto the alumina surface due to the amphiphilic character. Thus, from the resulting interaction, it is expected for the non-polar tail to remain exposed to the liquid, which confers to that region of interaction a reduced affinity to polar molecules, as water. By means of this mechanism, partial hydrophobicity can be attributed to alumina particles. These particles can, then, be attached irreversibly at the water-air interface, stabilising bubbles and giving rise to an ultrastable foam. The interaction of amino acids with $TiO_2$ particles was evaluated by Celani *et al.* [39], which prepared foams with this oxide.

In order to verify the hypothesis formulated, the evaluation of selected amino acid interaction with alumina in different pH values was investigated in this study by quantum mechanics simulations and experimental tests. These systems could enable the production of multiphasic macroporous refractory ceramics for thermal insulation, whose microstructures are tailored to reduce the thermal transportation phenomena responsible for energy losses in high temperature processes. Nonetheless, the production of foams with these features could also be applied in other sectors, as lightweight structural elements, scaffolds and catalyst supports, for example.

## 2. Materials and methods
### 2.1 Simulations

Quantum mechanics calculations for propionic acid, glycine, L-valine, L-isoleucine and L-leucine molecules (presented in Table 1) were carried out in order to study their structural properties at different pH. The electric dipole momentum expressed in Debye (D), the partial charge distribution along the molecules and the electronic density of neutral, anion, cation and zwitterion structures, were evaluated. Considering these results, the concentration of charges in different regions of the organic structures was clarified, making it possible to identify the presence of polar head and non-polar tails in these molecules, at different pH values.

The molecules were drawn with the help of Gabedit 2.5.0 [40] and were relaxed with semi empirical calculations with the MOPAC2016 software [41], using the Hamiltonian PM7 [42]. Then, the molecular geometries were optimised with the Hartree-Fock method (HF) [43], followed by another optimisation stage by Density Functional Theory (DFT) [44]. In both optimisation stages, the functional B3LYP [45] and the base Pople 3-21G [46] were used. Finally, the desired properties were assessed by DFT with the functional O3LYP [47] and the base Pople 6-31G(2p,2d) [48]. The optimisation step and the calculation of molecular properties were carried out in ORCA 4.1.1[49]. Water was selected as a solvent, and the Conductor-like Polarizable Continuum Model (CPCM) was used in the simulations [50]. During the property calculations, the results from the Mulliken population analysis were recorded [51], which was used to obtain the partial charge of each atom. The partial charge density surfaces were processed with Gabedit 2.5.0, using the van der Waals surface of the molecule as a drawing grid.

Table 1 Charged amino acids and carboxylic acid molecules studied in the mechano-quantum simulations.

| Molecule | Anion | Zwitterion | Cation |
|---|---|---|---|
| Glycine | 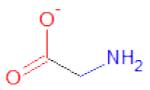 | 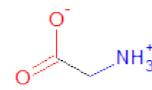 | 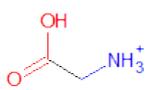 |
| Valine | 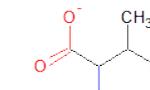 | 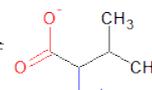 | 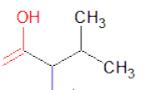 |
| Isoleucine | 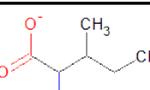 | 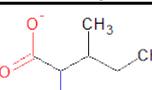 | 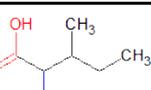 |
| Leucine | 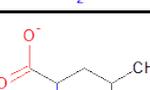 | 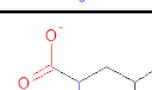 | 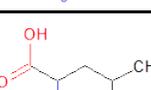 |
| Propionic acid | | 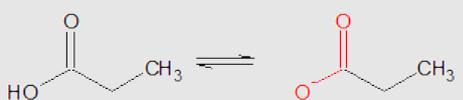 | |

**2.2 Experimental**

Aqueous suspensions with 35 vol% of alumina (CT3000SG, $d_{50} = 0.4$ μm, specific surface area = 7.5 $m^2g^{-1}$, Almatis, Germany) were prepared by deagglomerating the ceramic particles in distilled water and adjusting the pH system to 5 adding $HCl_{(aq)}$ at 1 $mol.L^{-1}$. Afterwards the suspension was transferred to polyethylene vessels, which contained high alumina grinding media (spheres with a diameter close to 10 mm), with a mass ratio of 2:1 for the grinding media and suspension, respectively. The ceramic suspensions were ball milled for 24 hours before being used in the experiments.

The selected amino acids were Glycine ACS, L-Valine, L-Isoleucine P.A. and L-Leucine P.A. (all analytical grade, by Synth, Brazil). Propionic acid (analytical grade, Sigma-Aldrich, Brazil), which is a molecule known to be used in the production of alumina stabilised foams in acid pH [34], was also selected. Their molecular structures, pK values and molar mass are listed in Table 2.

For some experiments, amino acids or carboxylic acid solutions were prepared with distilled water. The solution concentration was 0.15 $mol.L^{-1}$, which was close to the solubility limit of the tested amino acids in water at 25ºC. After preparing the solutions, their pH were adjusted to values close to pK listed in Table *2* with $HCl_{(aq)}$ or $NaOH_{(aq)}$, both at 1 $mol.L^{-1}$.

Table 2 Molecular structure, pK and molar mass of propionic acid and amino acids evaluated in this work. Data collected from [38].

| Molecule | Molecular structure and reduced formula | pK$_{cation}$ | pK$_{Zwitterion}$ | pK$_{anion}$ | Molar mass (g.mol$^{-1}$) |
|---|---|---|---|---|---|
| Propionic acid | 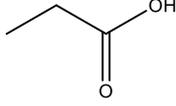 Propionic acid<br>$C_3H_6O_2$ | - | - | 4.7 | 74.08 |
| Glycine | 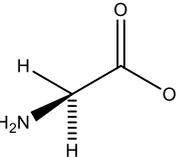 Glycine<br>$C_2H_5O_2N$ | 2.3 | 6.0 | 9.6 | 75.07 |
| L-Valine | 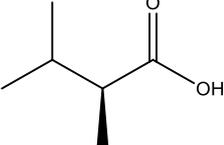 L-Valine<br>$C_5H_{11}NO_2$ | 2.3 | 6.0 | 9.6 | 117.15 |
| L-Isoleucine | 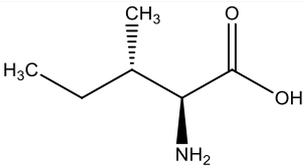 L-Isoleucine<br>$C_6H_{13}NO_2$ | 2.4 | 6.0 | 9.6 | 131.17 |
| L-Leucine | 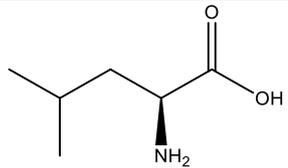 L-Leucine<br>$C_6H_{13}NO_2$ | 2.4 | 6.1 | 9.7 | 131.17 |

The foamability of the systems, *i.e.* their ability to entrap air, generating stable bubbles which resulted in a foam, was evaluated with the progressive addition of solutions containing amino acids or propionic acid. For that, a portion of 50g of the alumina suspension was transferred to a beaker. Following that, the suspension was kept under mechanical stirring by a dispersing blade with 38 mm of diameter, rotating at 500 rpm. Tiny amounts of solutions were added to the stirring suspension every 5 minutes, and the aspect of the systems were monitored by visual inspection in order to identify the formation of stable bubbles. Using this simple test, it was possible to identify the range of molecule concentration needed to generate foams.

The interaction of amino acids and propionic acid with $Al_2O_3$ in different pH values was evaluated by the indirect measurement of charge density formed on the particle surface. Suspensions containing 10 wt% of alumina, obtained after the dilution of previously ball milled suspensions, were prepared. Their pH was set to 5, 6 or 12 and kept at these values during the whole experiment. pH adjustments were carried out adding $HCl_{(aq)}$ or $NaOH_{(aq)}$, both at 1 mol.$L^{-1}$. The zeta potential ($\zeta$) of particles was measured as a function of the amount of molecules added. After a new addition of amino acids or propionic acid solutions, the system was kept under agitation for 10 minutes and zeta potential was then measured. The equipment ZA500 (Matec Applied Sciences, USA), which measures electrokinetic potential by the electroacoustic method, was used in these measurements. The equipment was calibrated with silica sol containing 10 vol% of $SiO_2$ (diluted Ludox TM40, Sigma-Aldrich, USA), and the tests were carried out at 25ºC. The pH of the suspensions was continually measured by a solid electrolyte-based pH electrode.

Contact angle measurements of amino acids or propionic acid solutions and alumina surface were carried out by the sessile drop technique in a goniometer ramé-hart 260 (ramé-hart Instruments, USA). Dense and polished alumina plates ($\rho = 3.96$ g.cm$^{-3}$) were used in this experiment. Before the measurements, alumina surfaces were submitted to a cleaning procedure consisting of immersion in the piranha solution (2:1 parts in volume of $H_2SO_4$ and $H_2O_2$) at 60ºC for 4 hours. Afterwards, the plates were thoroughly rinsed with distilled water, and then immersed in boiling distilled water for 1 hour. Finally, the alumina plates were rinsed with distilled water and kept in a desiccator until they were measured.

The contact angles were measured at 25ºC and, prior to the experiment, the camera was calibrated with measurements of deionized water drops over a Teflon® plate. Finally, the contact angle of amino acids and propionic acid were evaluated. For each molecule, the contact angles of three drops in different alumina surface regions were measured. The angles observed at the right and left extremities of the drop on the surface were recorded, and the values reported are the average of all measurements with their respective standard deviation.

The foam stability was assessed using equipment specially created for this purpose. This device comprises three humidity sensors (electrodes able to measure electric resistance in water), which are connected to an Arduino-based board (Blackboard UNO R3, Robocore, Brazil). Each sensor was placed at a different height of a polymeric cylinder with a volume close to 200 mL, which was kept closed with a polymeric lid. The local humidity was monitored as a function of time in three different regions of the fresh poured foam: bottom, middle and top. With the humidity progress, it was possible to follow the changes in the foam structure associated to the action of ageing phenomena, especially drainage, as the concentration of liquid at the bottom of a foam will decrease the electrical resistance, increasing the assessed conductivity value. The foam stability index (SI) of a foam region was obtained by the ratio between the resistance value $S_{(t=i)}$ assessed by the sensor at a time $t_i$, and the value $S_{(t=0)}$ measured at the beginning of the experiment, as expressed in the equation (3).

$$SI = \frac{S_{(t=i)}}{S_{(t=0)}} \tag{3}$$

If SI > 1 in equation (3), the amount of water is reduced in the region of measurement, which points to a drying process. On the other hand, if SI < 1, the amount of liquid phase is increased due to drainage (if it takes place at the bottom) or due to capillary flow resulting from liquid migration in a drying process (if it takes place at the middle or at the top of the foam). In this work, the measurements were carried out up to 100 hours.

## 3. Results and discussion

The electric dipole moment value indicates the polarisation of a molecule as it estimates the separation of opposite charges in the structure. As shown in Figure 1, neutral molecules of propionic acid, glycine and valine present smaller electric dipole moments when compared to isoleucine and leucine. The latter molecules are highly polar even in their neutral condition.

As expected, the molecules presenting charges in carboxyl or amine functions (anion, zwitterion or cations) are more polar than the neutral ones. For anions, the dipole moment was proportional to the size of the amino acid molecule, which can be related to the higher concentration of negative charge for the carboxyl, placed at the extremity of the molecule, accompanied by a partial reduction of charge intensity for amine and amino acid tails. For all cases, zwitterionic molecules were more polar because of the presence, in the same structure, of positive and negative charges at amine and carboxyl groups, respectively. Finally, a reduction in the electric dipole moment was observed for cation molecules. Additionally, the values were close and were not directly proportional to the molecular size.

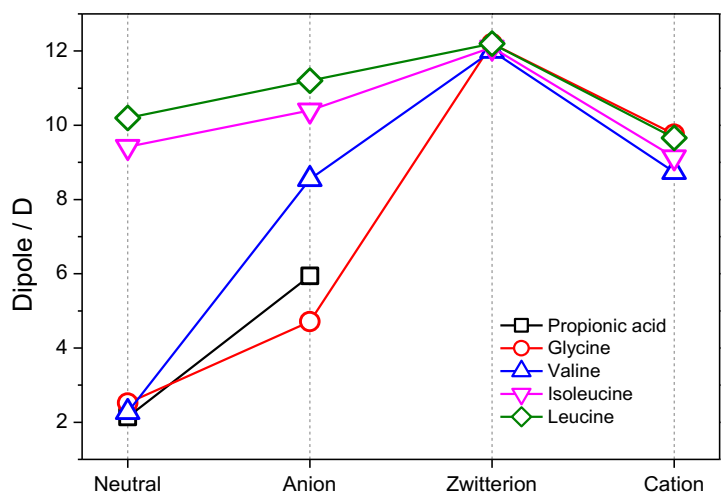

Figure 1 Calculated electric dipole moment of neutral, anionic, zwitterionic and cationic molecules.

The partial charge distributions for the molecules (Figure 2) highlights that most of the positive and negative charges are concentrated in carboxyl and amine functions, which are close to each other in a small region of these molecular structures. Thus, regardless of the pH, the amino acids studied here will present a polar head and a non-polar tail, which is not fully absent of charges. Non-polar tails of zwitterions and cations will present charge density superior than those calculated for anions. In these latter structures, negative charges are more concentrated in carboxyl and amine, increasing the hydrophobic character of the tail.

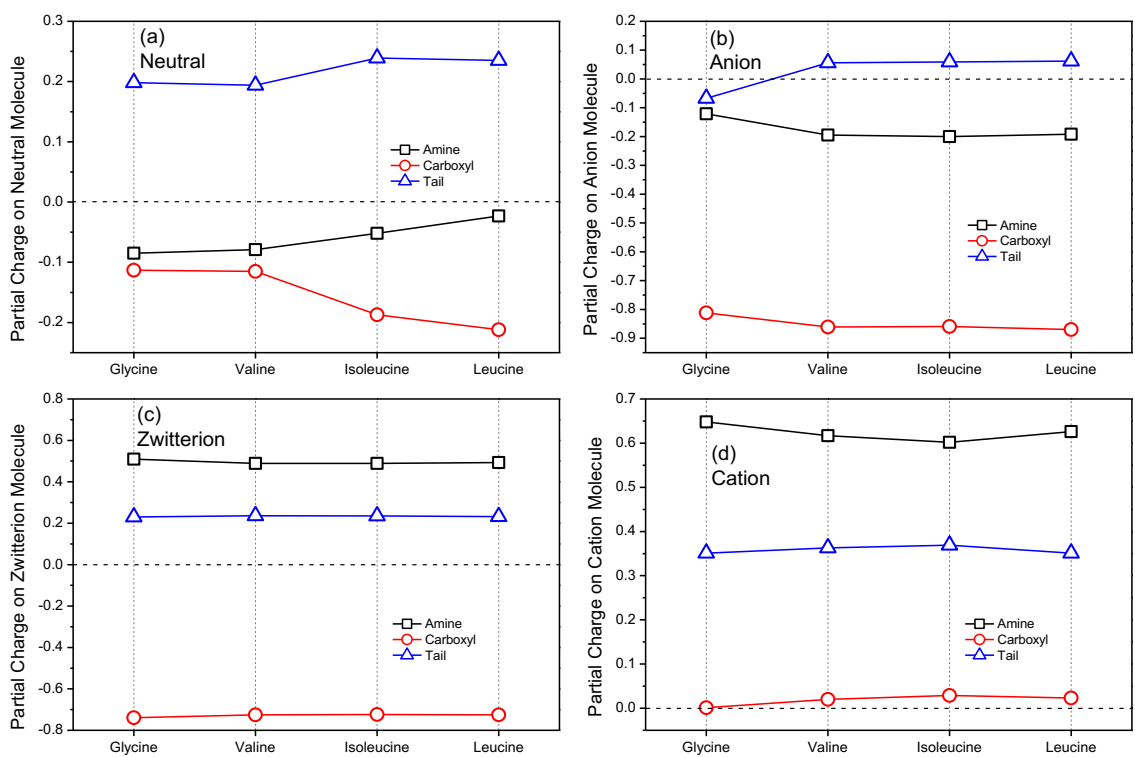

Figure 2 Partial charges distributed among amine, carboxyl and tail for neutral, anionic, cationic and zwitterionic amino acid molecules.

The results in Figure 2 point out a very small variation of partial charges at the tail of the amino acids. Thus, this charge should be distributed among all atoms of the tails, leading to the reduction in partial charge density for bigger structures, which reinforces their hydrophobic feature. Considering leucine and isoleucine molecules, the methyl position will also cause some displacement in charge distribution, especially for cationic structures. However, the partial charge values would be close to those observed for the other amino acids evaluated. The relationship between tail size and hydrophobicity for glycine, valine, isoleucine and leucine was consistent with trends pointed out by different hydrophobicity scales for these molecules. These scales were built based on thermodynamic calculations of the free energy changes related to the transfer of amino acids from a polar medium to a non-polar one [52–54]. In descending order, the tail hydrophobic character for the amino acids selected in their cationic, anionic or zwitterionic structures should be: leucine > isoleucine > valine > glycine. For these simulations, the values of partial charge, electric dipole moment and electronic density maps are available in the supplementary material (Table S1 and Figure S1).

The coupling between hydrophobicity and molecular size suggested by the simulations was also verified experimentally by the measurements of amino acid solutions contact angle on cleaned alumina surface (Figure 3). Leucine and isoleucine solutions presented higher contact angle then valine and glycine. Additionally, for all the cases, solutions whose pH were close to $pK_{Zwitterion}$ (see Table 2) had superior values of the contact angle. If the trend observed for the $\theta$ variation with pH for each amino acid is compared with their electric dipole moments (Figure 1), a relationship can be observed as the polarisation measurement seems to be coupled with the contact angle (an indirect measurement of hydrophobicity). Thus, for simple amino acids, the electric dipole moment could be used to forecast the behaviour of such molecules when in contact with hydrophilic surfaces. For all cases, amino acid solution drops presented a lower contact angle with the alumina surface when compared with distilled water ($\theta = 12°$) and propionic acid ($\theta = 45°$ at pH 5). Consequently, the amino acids evaluated in this study could result in partial hydrophobization of alumina surfaces, favouring the production of particle stabilised foams irrespective of the pH. Additionally, their adsorption energy at water-air interface would be higher than those expected for particles interacting with propionic acid.

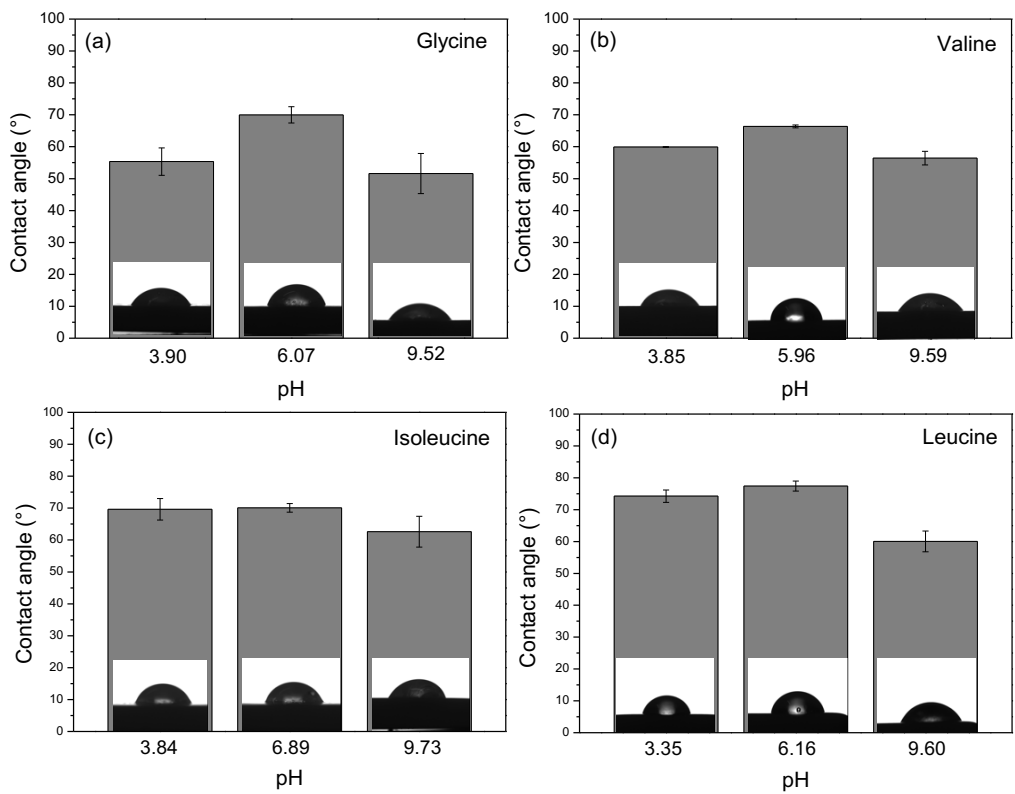

Figure 3 Contact angle of amino acid solutions (0.15 mol.L$^{-1}$ and different pH) on clean polycrystalline alumina surface at 25ºC. Contact angle of distilled water and propionic acid (pH = 4) on alumina were also evaluated, resulting in 12º and 45º, respectively.

It was possible to produce foams with amino acid solutions for different pH values, as seen in Figure 4. Foams were easily prepared when larger amino acid molecules (leucine and isoleucine) were used, followed by valine and glycine. In fact, a foam was not obtained when glycine was added to the suspension. The minimum amount of leucine and isoleucine needed to stabilise bubbles were close to the quantity of propionic acid required to foam a suspension at pH 5 (close to 5 µmol.m$^{-2}$, also verified by Gonzenbach *et al.* [34]). In addition, a minor quantity of amino acid in the zwitterionic structure was needed to produce a stable foam, which agrees with the quantum mechanics simulations and contact angle measurements. Besides the best foamability resulted by zwitterions, another advantage is that these structures are stable in the pH range close to neutral, which fulfils the Green Chemistry principles [37].

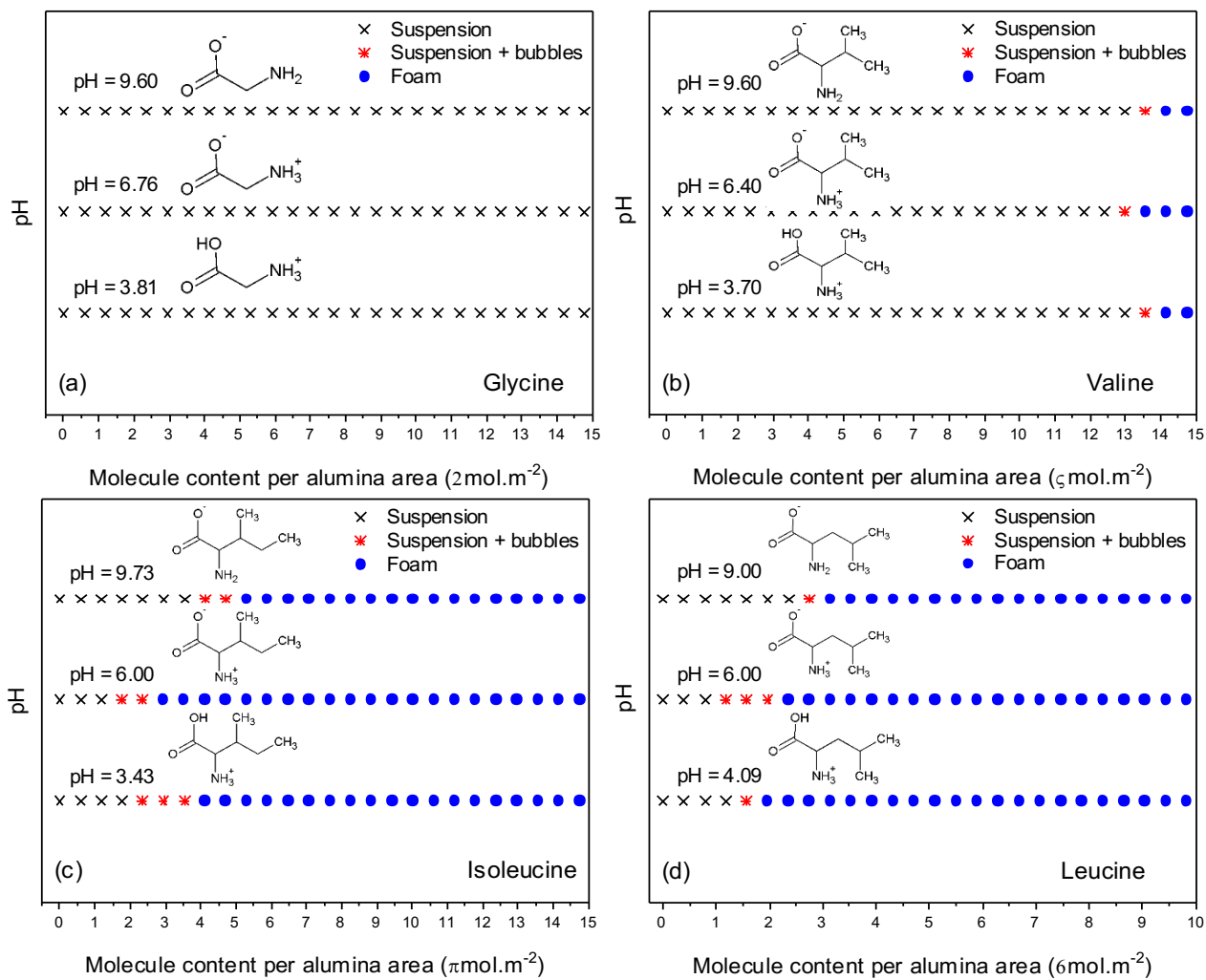

Figure 4 Empirical foamability maps for alumina suspensions containing 35 vol% of solids and increasing contents of amino acid solutions at 0.15 mol.L$^{-1}$, whose pH was adjusted to the values indicated in the graphs. For propionic acid, foams were obtained with 5 μmol per m$^2$ of alumina.

The alumina surface interaction with leucine, isoleucine and propionic acid was assessed by zeta potential measurements (Figure 5). The reduction of ζ for suspensions with propionic acid, kept at pH 5 and 6, was observed, whereas for pH 12, the electrokinetic potential did not change. On the other hand, variations of zeta potential were observed for all pH values evaluated when isoleucine or leucine were added to alumina suspensions.

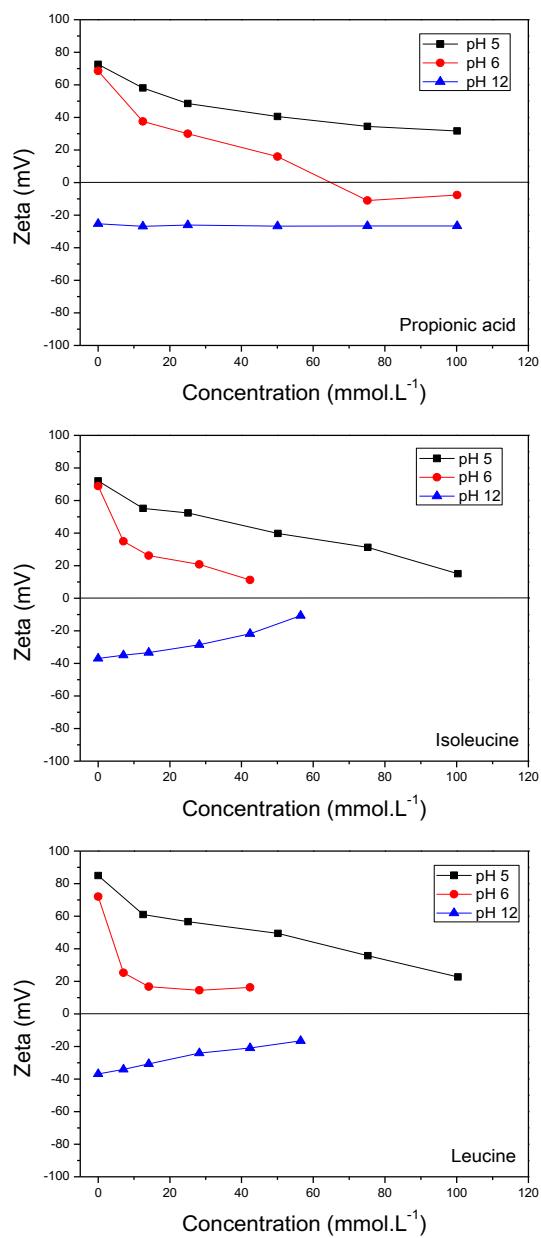

Figure 5 Zeta potential changes for alumina particles as a function of the increasing amount of propionic acid, isoleucine or leucine molecules. The pH was kept constant during all the experiments.

The reduction in the absolute value of ζ may take place due to a charge screening effect on the oxide surface, caused by the adsorption of charged molecules [55]. In this case, the molecule polar head may be directed towards the particle surface, which presents positive or negative charge density, whereas the non-polar tail remains exposed to the aqueous media. Thus, the results presented in Figure 5 suggest that leucine and isoleucine molecules adsorbed on alumina surfaces for all pH values evaluated. For propionic acid at pH 6, an inversion was observed on the electrokinetic potential between 45 and 65 mmol.L$^{-1}$, most likely due to the arrangement of a second layer of adsorbed molecules, whose orientation was inversed and the polar head, formed by carboxyl negatively charged, was exposed to the aqueous side. This effect was identified for similar systems by Megias-Algacil *et al.* [56].

Particle stabilised foams are called ultrastable because of the long period of time in which they remain without undergoing ageing phenomena. The evolution of drainage was assessed in this study by measurements of electric conductivity in regions of fresh foam columns (bottom, middle and top). The results shown in Figure 6(a) point out that foams produced with 10 μmol.m$^{-2}$ of propionic acid experienced structural changes due to drainage. For this system, the sensors placed at the bottom region and at the middle registered a reduction in the foam stability index because of the increasing amount of liquid in these regions. After 15 hours approximately, the sensor placed at the top lost contact with the foam column, and the test was stopped. When the propionic acid amount was twofold [Figure 6(b)], the produced foam presented superior lifetime, and drainage was observed at the beginning. Sequentially, the curve profiles suggest the drying of the foam due to evaporation of the liquid. For the foam containing 10 μmol.m$^{-2}$ of isoleucine, a similar process of drying was identified [Figure 6(c)]. When the amount of this amino acid was increased [Figure 6(d)], the foam did not present any stability index changes for the first 100 hours. The same results were observed for all foams prepared with leucine and indicated that foams produced with partially hydrophobic amino acid-based aluminas were more stable than those produced with propionic acid, for the conditions adopted in this study.

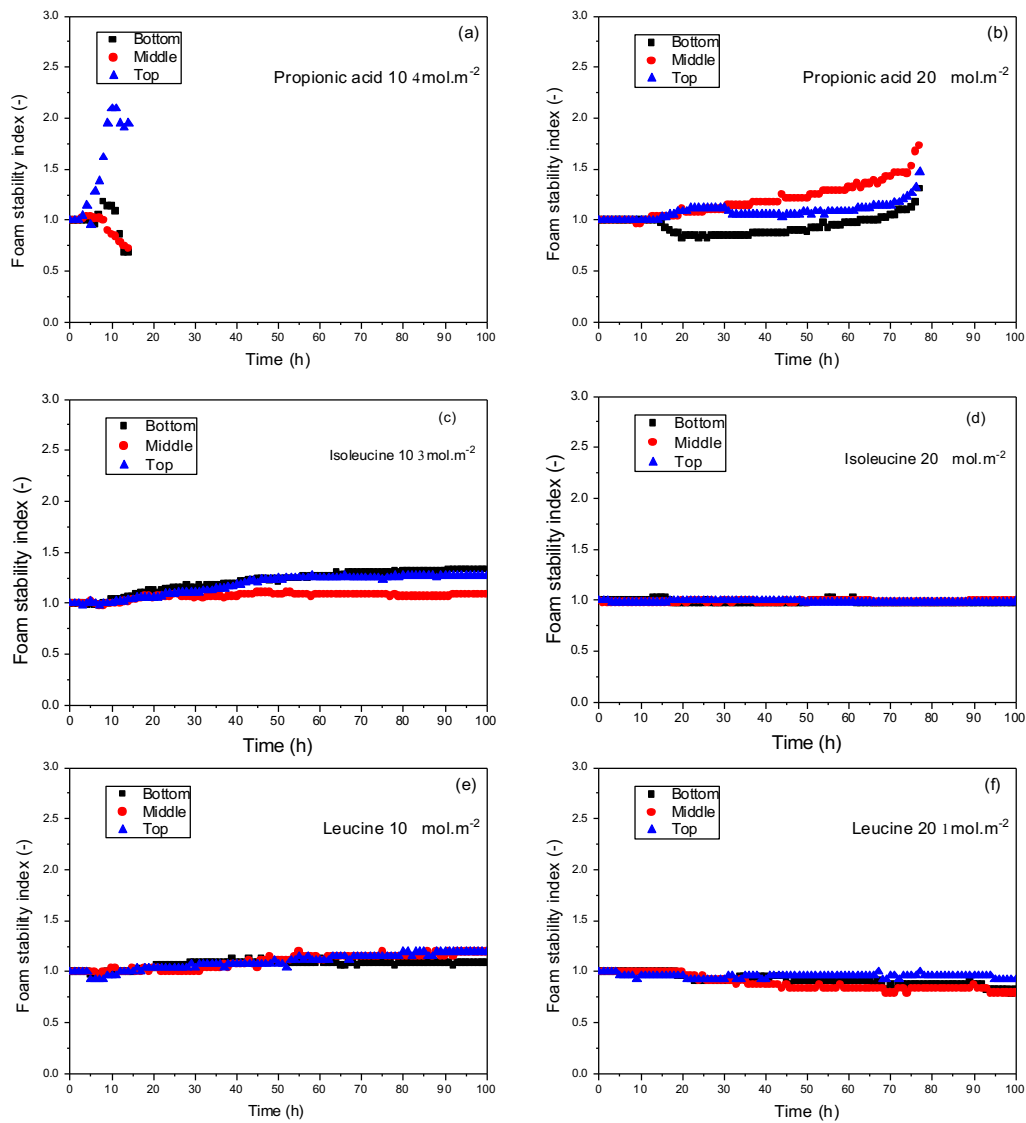

Figure 6 Evaluation of alumina foam stability with time at 25°C for systems prepared with alumina suspensions at pH close to 5, containing 10 or 20 μmol.m$^{-2}$ of propionic acid, isoleucine or leucine.

Most likely due to the high amount of entrapped air and the disjunction pression acting between the bubbles, both favouring a jamming transition [20], foams produced with partially hydrophobic aluminas can present high yield stress (see Figure 7). This feature indicates that these foams can be used as feedstocks for additive manufacturing techniques, in a similar way as Muth *et al.* [57] did with propionic acid-based alumina ones. Figure 7(c) presents the microstructure of this foam after firing at 1600°C for 5 hours, highlighting the elevated number of small and closed pores in this sample. Considering the likelihood of stabilisation of the foams discussed in this work to a broader pH range, they could be used for the production of multiphasic compositions, an issue that will be addressed in forthcoming work.

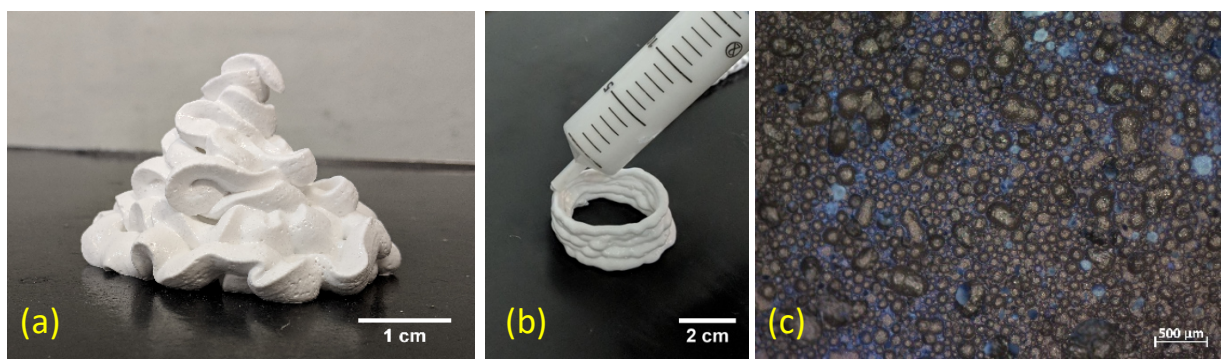

Figure 7 (a,b) Fresh alumina stabilised foams prepared with alumina suspension and 20 µmol.m$^{-2}$ of L-leucine and (c) microstructure of this foam after firing at 1600°C for 5h, as seen by a stereo microscope. The surface was previously painted with blue ink to ease the identification of pores.

4. **Conclusions**

Glycine, L-valine, L-isoleucine and L-leucine were evaluated as potential molecules for partial hydrophobization of alumina particles. Considering the quantum mechanics simulation results, due to the existence of carboxyl and amine functional groups, these molecules present charge concentration in a small region of their structures, regardless of the pH, giving rise to a polar head and a non-polar tail. The charge concentration at the non-polar tail is similar for all amino acids studied, which indicates that the hydrophobic character increases with the molecular size. Thus, the hydrophobicity trend decreases as follows: leucine > isoleucine > valine > glycine, which was confirmed by experiments.

L-Leucine and L-isoleucine solutions displayed higher contact angles on a clean alumina surface when compared to L-valine and glycine solutions. Besides that, all evaluated amino acids wetted less the alumina surface when compared with propionic acid, for all pH ranges evaluated. However, among the likely amino acid species to be formed, zwitterions were more attractive to be used in the partial hydrophobization of alumina, as these structures present higher electric dipole moment, contact angle and interaction with the oxide. These aspects indicated that zwitterions can be more efficient for the hydrophobization of alumina surfaces and consequently, formation of stable foams. Indeed, ultrastable foams produced with alumina and leucine or isoleucine were much more stable than those prepared with propionic acid. The obtained foams can be easily extruded, which opens up the possibility of using them as feedstocks for additive manufacturing. Besides that, the results suggested that ceramic particles with distinct charge densities could be added to the foams, without destabilising them. Thus, foams stabilised with amino acid-based partially hydrophobic aluminas could comprise multiphasic macroporous ceramics whose properties could be tailored for different applications, such as thermal insulation for high temperature application or lightweight hierarchical structures.

## 5. Acknowledgements

This study was financed in part by the Coordenação de Aperfeiçoamento de Pessoal de Nível Superior - Brasil (CAPES) - Finance Code 001. The authors are also thankful to FIRE (Federation for International Refractory Research and Education),

FAPESP (Fundação de Amparo à Pesquisa do Estado de São Paulo, Grant Number 2018/09761-8) and Cloud UFSCar for all the support for this research.

**Supplementary information**

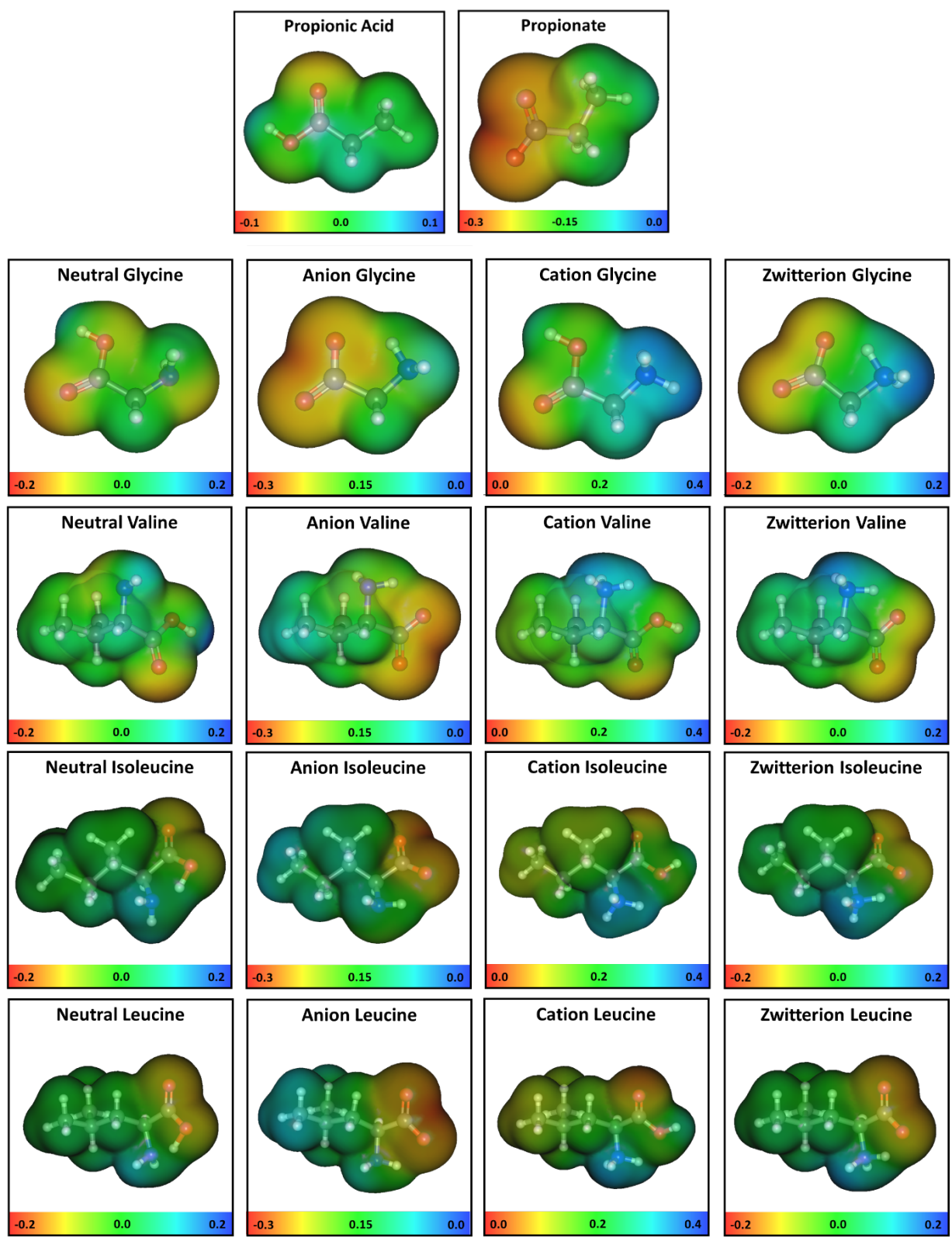

Figure S 1 Electric density surfaces (scales corresponding to partial charges) for the molecules evaluated in this work

Table S 1 Partial charges and electric dipole moment (in D) values for the molecules evaluated in this work

|  |  | Amine | Carboxyl | Tail | Total | Dipole (D) |
|---|---|---|---|---|---|---|
| **Propionic acid** | Neutral | - | -0.088 | +0.088 | 0.00 | 2.14 |
|  | Anion | - | -0.866 | -0.134 | -1.00 | 5.94 |
| **Glycine** | Neutral | -0.085 | -0.113 | +0.198 | 0.00 | 2.52 |
|  | Anion | -0.121 | -0.812 | -0.067 | -1.00 | 4.71 |
|  | Cation | +0.648 | +0.001 | +0.351 | +1.00 | 9.76 |
|  | Zwitterion | +0.509 | -0.739 | +0.230 | 0.00 | 12.2 |
| **Valine** | Neutral | -0.079 | -0.115 | +0.194 | 0.00 | 2.27 |
|  | Anion | -0.195 | -0.861 | +0.056 | -1.00 | 8.55 |
|  | Cation | +0.617 | +0.020 | +0.363 | +1.00 | 8.74 |
|  | Zwitterion | +0.489 | -0.725 | +0.236 | 0.00 | 12.0 |
| **Isoleucine** | Neutral | -0.052 | -+0.187 | +0.239 | 0.00 | 9.42 |
|  | Anion | -0.200 | -0.859 | +0.059 | -1.00 | 10.4 |
|  | Cation | +0.602 | +0.029 | +0.369 | +1.00 | 9.14 |
|  | Zwitterion | +0.489 | -0.724 | +0.235 | 0.00 | 12.1 |
| **Leucine** | Neutral | -0.023 | -0.212 | +0.235 | 0.00 | 10.2 |
|  | Anion | -0.192 | -0.870 | +0.062 | -1.00 | 11.2 |
|  | Cation | +0.626 | +0.023 | +0.351 | +1.00 | 9.66 |
|  | Zwitterion | +0.493 | -0.725 | +0.232 | 0.00 | 12.2 |